\documentclass[12pt]{article}
\usepackage{graphicx}
\usepackage{MnSymbol}
\usepackage{bbm}
\usepackage{amsfonts}
\usepackage{authblk}

\oddsidemargin=\evensidemargin
\def\postwomu{\mskip 2mu}
\def\Sc{Schr\"odinger}
\def\tH{'t$\postwomu$Hooft}
\newcommand{\be}{\begin{equation}}\newcommand{\ee}{\end{equation}}

\newcommand{\labeld}[1]{ }

\def\Eqref#1{Eq.~(\ref{#1})}
\def\PsiUP{\psi_\mathrm{up}\otimes\psi_\mathrm{bath}}

\def\one{{\mathbbm 1}}

\def\assume{I assume familiarity with the \Sc\ cat paradox.}
\def\donotmodel{The irreversible ``registration'' of the result of a measurement by the observer has been studied in many contexts. For example, in \cite{lsmodel} the ``measurement'' is accompanied by the bath's (not the same as the bath in the current \Eqref{e.010}) changing in an irreversible fashion. Other models of measurement (e.g., \cite{bona, green}) show the same feature. As a result, our considerations in the present article do not pursue the registration issue once the observer is coupled to the system, that coupling taking place (in our forthcoming example) at 0.15 time units.}

\begin{document}

\title{\vskip -40pt
\bf Need for ``special" states in a deterministic theory of quantum mechanics}
\author{L. S. Schulman}
\affil{Clarkson University, Potsdam, New York 13699-5820, USA\\
{\small email: {\tt schulman$<$at$>$clarkson.edu}}}

\date{\today}

\maketitle

\begin{abstract}
There are several theories or processes which may underlie quantum mechanics and make it deterministic. Some references are given in the main text. Any such theory, plus a number of reasonable assumptions, implies the existence of what I have called ``special" states. The assumptions are conservation laws, obedience (up to a point) of \Sc's equation, no hidden degrees of freedom and a single world, in the sense of the many worlds interpretation (the last one a consequence of any deterministic theory). This article also, for clarity, gives an example of a ``special" state. There is an experimental test of the ``special" state theory.
\end{abstract}

\section{Introduction\label{s.intro}}

Determinism is a loophole in Bell's \cite{bell} ideas, which he was aware of. I unwittingly exploited it in 1984 \cite{lsdefpla} with what I will call the ``special" state theory of measurement. (In Sec.\ \ref{s.special} an example of a ``special" state is given.) The present article reports new motivation for ``special" states.

There have been quite a few attempts to find an underlying process that would make the \Sc\ equation deterministic. I am \textit{not} referring to Bohm's interpretation \cite{bohm} or that of his followers. Rather I have in mind those theories which would restore determinism, such as (\textit{not exclusively}) those of \tH\ \cite{thooft1,thooft2}, Palmer \cite{palmer}, De la Pena Auerbach and Cetto \cite{delapena}, Cavalleri et al\@. \cite{cavalleri}, Cufaro-Petroni and Vigier \cite{cufaro} and Marshall \cite{marshall}. For at least some of these the \Sc\ equation is an approximation---a good approximation, but an approximation nevertheless. There has also been discussion about the consequences of determinism \cite{thooft1,palmer,hossenfelder,brans,hall}.

There is an experimental test of the ``special" state theory, which, if successful, would lend credence to some of the theories advanced. If negative, it would be challenging to maintain determinism.

\section{``Special" states \label{s.special}}

Most of this section is review. It may be skipped by those familiar with the kind of ``special" states that I have in mind.

\medskip

We take, as an example of a ``special" state, a spin, initially pointing  in the positive $z$ direction with a 50\% probability of overturning at some given time, say at $0.15$ (since all is determined the time of observation is also fixed).  Moreover, we don't deal with ``registration" of the measurement; that will be accomplished by additional degrees of freedom.\footnote{\donotmodel} The Hamiltonian is
\be
 H=\frac\varepsilon2     \left( 1+\sigma_z  \right)
   + \omega a^\dagger a
   +\beta\sigma_x (a^\dagger + a)
\,.
\label{e.010}
\ee
\labeld{e.010}
The Pauli matrices $\sigma_x$ and $\sigma_z$ are the operators for the 2-state spin system, $a$ and $a^\dagger$ are the boson operators and $\varepsilon$, $\beta$ and $\omega$ are parameters.

\textit{``Special" states} are particular initial conditions of the bath such that the \textit{microscopic} final state of the spin is (either) \textit{all} \textsf{up}, $\left({e^{i\phi_1}\atop0}\right)$, or \textit{all} \textsf{down}, $\left({0\atop e^{i\phi_2}}\right)$\@. ``Final'' refers to the time of measurement, namely when (even) more degrees of freedom are involved (we use parameter values $\epsilon=0.5$, $\omega=0.1$, $\beta=0.6$ and a time of 0.15).

\begin{figure}[ht]
\centerline{
\includegraphics[width=.4\textwidth]{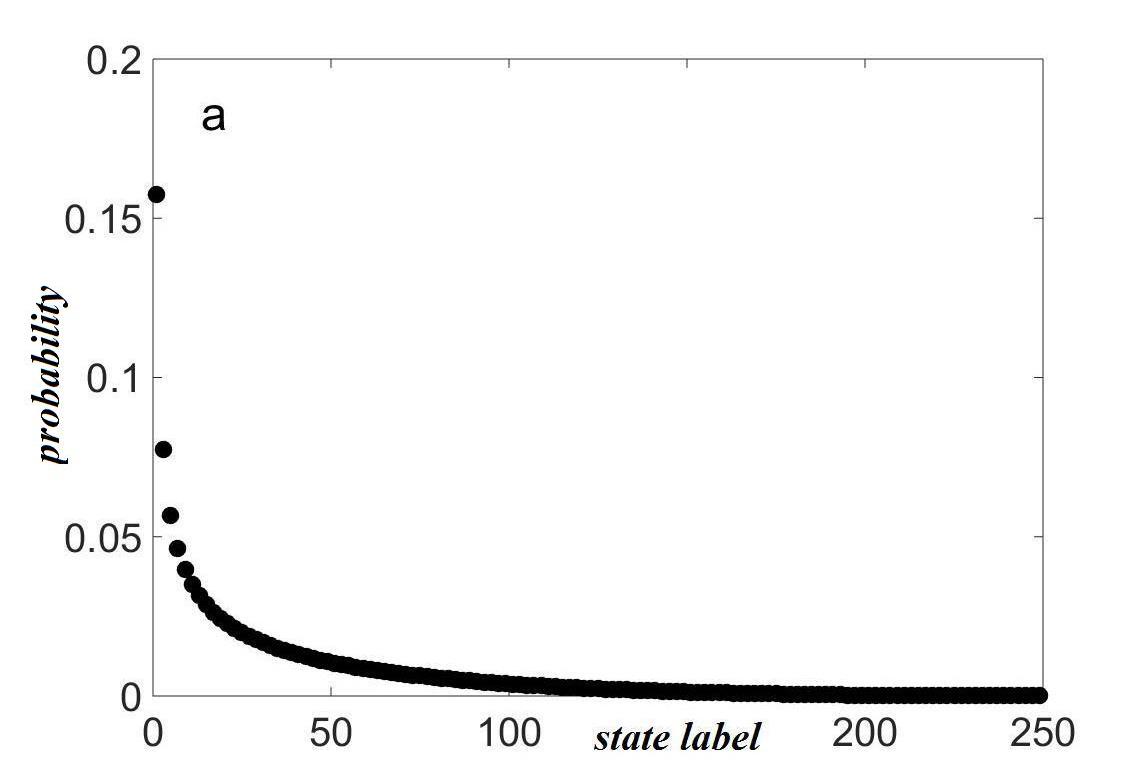}
\includegraphics[width=.4\textwidth]{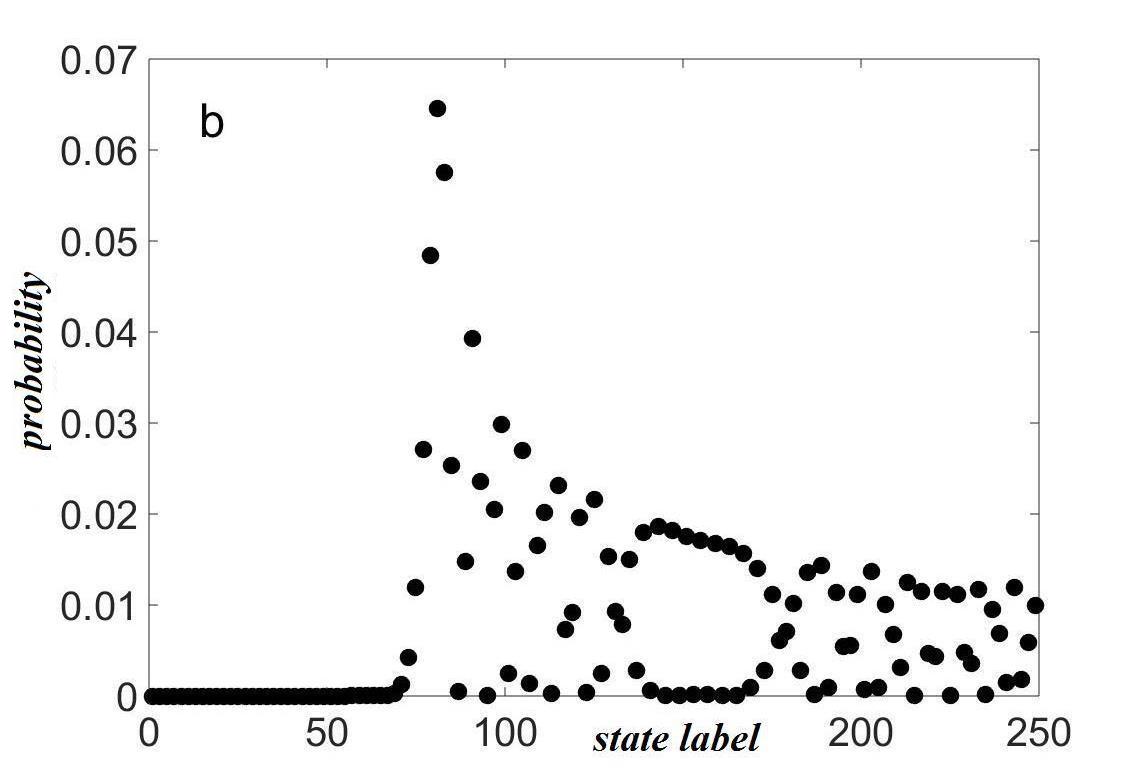}
}
\caption{``Special" \textit{time-}0 oscillator states. Figure (\textbf{a}) shows the (initial) probability of excitation of oscillator states that contribute to the non-decay state. Only shown are even states, since there is total amplitude zero for the odd states. Phases of the states are not shown, but are also fixed by the non-decay condition. In Figure (\textbf{b}) are shown the probabilities for the state that decays; in this case (and for the same reason) only even oscillator states are shown. As in image \textbf{a}, the phases, though not shown are crucial to the ``special'' nature of the state. \label{f.1}\labeld{f.1}}
\end{figure}

The system begins in all \textsf{up} and ordinarily at time 0.15 has probability of half up, half down. As indicated, that is \textit{not} the case for these ``special" initial conditions. If the probabilities are as in Fig.\ \ref{f.1}a---with fixed phases (not shown)---then the system will be found in an \textsf{up} state. If the initial state is as shown in Fig.\ \ref{f.1}b (again with particular phases, not shown) then the system will be found in a \textsf{down} state.

There are three points to be raised: the first is what about residual amplitudes? The amplitude for (say) the \textsf{up} state is not perfect and for the given cutoff of the bosons at 250 is about $10^{-4}$; the same is true for the state which is ``fully" decayed. The second question has to do with \Sc's cat. And the third issue is how do you find these states?

Now $10^{-4}$ is a big number, especially since the final state of one interaction is the initial state for the next. I could improve that number if I had better computer power, but I doubt if it could be zero. But it doesn't have to be zero! It only needs to be accurate as far as the \Sc\ equation has been checked. And I don't think it has been checked to $10^{-12}$ (which I am reasonably confident I could get the discrepancy down to).

The second issue I mentioned is, what about \Sc\ cats? The (possibly) decaying spin could be the determinant of whether the poison is released.\footnote{\assume} With ``special" states the cat is either alive or dead. It should be noticed that there are only what \tH\ calls ``ontological" states \cite{thooft1,thooft2}. I believe ``special" and ``ontological" have the same meaning in this case.

Finally there is question of how ``special" states are found. You can define a projection operator (cf.\ \cite{lstimebook}) on the spin: $P\equiv \left(\psi_\mathrm{up} \psi_\mathrm{up}^\dagger \right)\otimes\one_\mathrm{boson~bath}$\@. Using this operator, the probability of being all \textsf{up} at time $t$ is
\be
\Pr(\hbox{\textsf{up}})=\langle\PsiUP|U^\dagger P U|\PsiUP\rangle
=\langle\PsiUP|PU^\dagger PP UP|\PsiUP\rangle
\,,
\ee
with $U\equiv \exp\left(-iHt/\hbar\right)$ and where $PP=P$ is used. Defining $A\equiv PUP$ and using $P^\dagger=P$, we have $\Pr(\hbox{\textsf{up}})=\langle\PsiUP|A^\dagger A|\PsiUP\rangle$\@. Defining $B\equiv A^\dagger A$, it follows that the issue of whether any initial state (of the bath) can lead to a measurement of \textsf{up}, using purely unitary time evolution is the matter of whether $B$ has eigenvalues equal to one. For any fully decayed states you must have an eigenvalue (of $B$) be zero. (Of course $A$ and $B$ are functions of $t$, since $U$ is.)

\smallskip

\noindent\textsf{Remark:~~} It also is true that the number of decay states and non-decay (``special") states is roughly equal at time 0.15.

\smallskip

This is the main idea of the ``special" state theory: no macroscopic superpositions because of particular initial conditions.  There is also no entanglement. At time-0.15 the spin state is wholly in one state or the other.

\section{Determinism implies ``special" states\label{s.determinism}\labeld{s.determinism}}

The title of this section needs a bit of enhancement: you need a few more concessions to reality. Besides determinism you need conservation laws and \Sc's equation, at least to the extent that it's been checked. You also need have no coordinates (degrees of freedom) invisible to the \Sc\ equation. It is understood that there is just one world. These rules, together with determinism, imply ``special" states.

You start with a wave function describing some state, say a spin in a Stern-Gerlach experiment. Then it \textit{must} go to some particular outcome, say spin \textsf{up}. Presumably there were involved other coordinates (such as the bosons in the above example) that fixed its outcome. The final state is definite. But the \Sc\ equation holds also. Therefore it could only have evolved to that final state. How can that be? There must have been a coordination of degrees of freedom on the initial state that forced it to its final form. That is, there must have been a ``special" state.

Note that if the particle is given another degree of freedom of the right sort the above argument does not hold. Thus if the particle is secretly (say) spin up (i.e., $J_z=1$ in a Stern-Gerlach experiment) then it can avoid having other degrees of freedom involved. Since the nature of the experiment ($J_z$ is measured, as opposed to say $J_x$) is determined this is legal. However, without this degree of freedom (presumably invisible to the \Sc\ equation) the argument for special states is valid.

\section{Experiment\label{s.experiment}\labeld{s.experiment}}

Finally, there is the matter of experiment. In \cite{lsexperiment} and \cite{lssource} we have described in detail experimental tests of the ``special" state theory. An example is the double Stern-Gerlach experiment (\cite{phipps,frisch,majorana,stern}) which requires the detection of a magnetic field of $5\times10^{-8}$ tesla in an environment of half a tesla, a challenging experiment. A firm absence of the small magnetic field would in my opinion spell the end of efforts to find a deterministic theory with no additional degrees of freedom.

\section{Conclusions\label{s.conclusions}\labeld{s.conclusions}}

You don't have to believe in any of the deterministic theories to reach the conclusion that ``special" states are needed in any theory which is deterministic, goes from one ``special" state into another, satisfies \Sc's equations (as far as has been measured), has a single world, does not involve degrees of freedom invisible to the \Sc\ equation and satisfies conservation laws. You only have to believe that it's possible.

Three points are worth mentioning. First---and this is new---you don't need to eliminate ``incorrect" choices (by ``special" states) at the level of (say) $10^{-12}$, since the \Sc\ equation has not been checked at that level. Second, there is an experimental test of the special state theory. Failure would eliminate deterministic theories (or leave people struggling for an explanation), while success would encourage attempts to find deterministic theories. Third, it may be that \tH\ is right, and one should look to extremely small times and distances for theoretical support for determinism. However, given the fragmentary understanding of events at $10^{-17}\;$cm I'd be reluctant to make predictions about what happens at $10^{-33}\;$cm.


\end{document}